\documentstyle[11pt,aaspp4,epsfig,psfig]{article}     

\def\ltsima{$\; \buildrel < \over \sim \;$}
\def\lsim{\lower.5ex\hbox{\ltsima}}
\def\gtsima{$\; \buildrel > \over \sim \;$}
\def\gsim{\lower.5ex\hbox{\gtsima}}

\begin{document}

\title{\bf 
RECYCLING NEUTRON STARS TO ULTRA SHORT PERIODS: A STATISTICAL ANALYSIS
OF THEIR EVOLUTION IN THE $\mu$ - $P$ PLANE.}

\author{Andrea Possenti\altaffilmark{1},
Monica Colpi\altaffilmark{2},
Ulrich Geppert\altaffilmark{3},
Luciano Burderi\altaffilmark{4},
Nichi D'Amico\altaffilmark{5}}

\altaffiltext{1}{Dipartimento di Astronomia dell'Universita', via Ranzani 1,
40127 Bologna, Italy}
\altaffiltext{2}{Dipartimento di Fisica dell'Universita', via Celoria 16,
20133 Milano, Italy}
\altaffiltext{3}{Astrophysikalisches Institut Potsdam, An der Sternwarte 16, 
14482 Potsdam, Germany}
\altaffiltext{4}{Osservatorio Astronomico di Monteporzio, via Frascati 33,
00044 Roma, Italy}
\altaffiltext{5}{Osservatorio Astronomico di Bologna, via Ranzani 1,
40127 Bologna, Italy}
\vskip 14pt
\centerline {\it To Appear in ApJS}

\begin{abstract}
In this paper we investigate the statistical evolution of magnetic neutron 
stars, recycled in binary systems, simulating synthetic populations. 
To bracket uncertainties, we consider a soft (FP) and a stiff (PS) equation
of state (EoS) for nuclear matter and explore the
hypothesis that the magnetic field is confined in the stellar crust.
We follow the magneto-rotational evolution within a simple recycling 
scenario. The decay of the magnetic field is modelled imposing at the
crust--core boundary either complete field expulsion by the superconducting 
core or advection and freezing in a very highly conducting transition shell.

Irrespective to the details of the physical models  
we find the presence of a tail in the period distribution of the
synthetic populations at periods shorter than 1.558 ms, the minimum
detected so far. For the soft EoS, and independent of the details of the 
magnetic field evolution, the recycling gives rise to a spin distribution
which is increasing  monotonically toward short periods and a  clear 
``barrier'' forms at the minimum period for the onset of mass shedding
($\simeq 0.7$ms). 
For the stiff EoS the distribution is flatter displaying a broad maximum 
about 2-4 ms. On the other hand, if in low mass binaries the
neutron stars experience   a progressive decrease of the mass accretion
rate 
(due to transient behaviour and/or the quenching of accretion), the
magnetospheric propeller produces (together with the magnetic dipole losses) 
an overall depletion of neutron  stars in the millisecond region of the
$\mu$-$P$ plane.
 
The estimated fraction of neutron stars spinning close to their 
shedding limit over the  millisecond pulsar population 
is found to be significant. Crustal magnetic field decay models 
predict also the existence of massive rapidly spinning neutron stars 
with very low magnetic moment $\mu<10^{25.8}$ G cm$^3.$

\end{abstract}


\section{\bf Introduction}

PSR1937+21, at present, is the neutron star having the shortest
rotational period $P_{min}=1.558$ ms ever detected: It is a millisecond radio
pulsar (MSP), the fastest among the $\sim 30$ discovered in the Galactic
field  with weak magnetic field and age $>10^9$ yr. Despite its apparent 
smallness, $P_{min}$ is not a critical period for NS rotation: NSs can indeed 
spin faster, as indicated by Cook, Shapiro \& Teukolsky (1994).  
These authors showed
that during recycling  a bare non magnetic NS can attain periods as short as $P_{min}$ 
with modest values of  the rotational to potential energy ratio and modest values of
the mass accreted ($< 0.1~{\rm M_{\odot}}$), for all known
equations of state (EoSs) of the nuclear matter. The period $P_{min}$ is 
longer than the limiting period, $P_{sh}$, below which the star becomes 
unstable to mass shedding at its equator, irrespective to the EoS.
It is clear that only the discovery of an ultra short period MSP
(i.e. a weak field pulsar with $P$ close to $P_{sh}$) would allow to set 
limits on the proposed EoSs; as an example, a very soft EoS permits
pulsations at $\sim 0.7$ ms.
 
Burderi, Possenti, Colpi, Di Salvo \& D'Amico (1999) noticed that the recycling  of
magnetic NSs to $P<10$ ms depends sensitively on the mass transferred to the 
compact object and crucially on the evolution of the surface magnetic field. 
In particular, a slowly decaying  magnetic field  favours
only the re-acceleration of a NS  to $P\gsim 5 \div 10$ ms, while halving
the request of mass accreted  with respect to the case of a nonmagnetic star.
Conversely, if a rapid and substantial decay 
of the surface magnetic field takes place, a NS can attain $P<P_{min}.$
In this case, the minimum mass load 
necessary to approach $P_{sh}$ is of $\sim 0.3~{\rm M_{\odot}}$: 
it was calculated using the softest EoS (having $P_{sh}=0.7$ ms) and
introducing the most favourable {\it ad hoc} evolution for the decay of the 
magnetic field.   In this paper we wish to study in details those processes
which interfere positively to spin up a NS up to $P_{sh}$ 
in a LMB. Among them, the most important are the evolution of the magnetic 
moment $\mu,$ and the evolution of the mass transfer.

At present, there is no satisfactory theory for the origin of the magnetic
field and consequently several avenues of field decay are possible: 
The field can either be a fossil remnant from the progenitor star or be 
generated soon after the formation of the NS. Thus, an unanswered question is 
where the bulk of the magnetic energy is stored (Bhattacharya \& 
Srinivasan 1991). If it resides mainly in the core (as in the case of a fossil
field), the evolution of $\mu$ may be driven by variations in the rotational
state of the NS, which affects the motion of the superfluid 
vortices and, in turn, of the fluxoids, carriers of the magnetic energy 
(Srinivasan {\it et. al} 1990; Ding, Chen \& Cau 1993; Miri \& Bhattacharya 
1994; Ruderman 1998). If the magnetic energy is generated and confined in 
the thin crust of the NS (Blandford, Hernquist \& Appelgate 1983;
Urpin, Levshakov \& Yakovlev 1986), the surface magnetic field decay is
guided  by the ohmic diffusion and the dissipation of electric currents 
(Sang \& Chanmugam 1987: see $\S~3$ for an updated set of references). 
The EoS valid for the core determines the thickness of the crust as well as 
the cooling history of the NS which both affect strongly the magnetic field 
decay. In the case of a crust field however, magnetic evolution is not 
univocal, depending sensitively  on the properties of the nuclear matter 
at the crust--core boundary, where superfluid (SF) and superconductive (SC) 
phase transitions may occur. As crustal matter is progressively assimilated 
into the core due to accretion, the magnetic field at the crust-core boundary 
can either be expelled (Urpin, Geppert \& Konenkov 1998) or advected into 
the SC core where it no longer decays (Konar \& Bhattacharya 1998). Thus, 
at the endpoint of evolution, the NS can either become non magnetic (on the 
long term), or preserve a relic field intense enough to shine again as MSP.
These avenues are possible and the extent of the decay is intimately 
connected with the transfer rate and amount of matter accreted as heat 
released in the crust rises the electron and ion resistivity, accelerating 
the dissipation of current. Thus, field and mass transfer evolution couples 
intimately during the lifetime of the NS in a LMB.

Within the recycling scenario (Alpar et al. 1982; Bhattacharya 1995) all schemes
suggested 
for the origin and the evolution of $\mu$ allow accreting NSs to reduce 
their magnetic moment to the values that are characteristic of MSPs
(Urpin, Geppert \& Konenkov 1998) but core models requires rather
extreme conditions on the impurity content of crustal matter 
(Konar \& Bhattacharya 1999). Only through a statistical approach 
it is possible to establish how efficient the recycling process is in 
spinning NSs to ultra short periods. This has been recognized first by 
Possenti, Colpi, D'Amico \& Burderi (1998, paper I) who carried on a 
statistical analysis of NSs in the millisecond and submillisecond interval, 
using a population synthesis model. Their aim was to determine whether there
might exist feasible conditions for spinning a significant number of NSs to 
$P<P_{min}$, during recycling in LMBs. They found first evidence of a tail 
in the distribution of MSPs at periods shorter than those detected so far. 
This 
possibility was met within empirical
models that involve  the screening by the accreting matter treated
as a pure diamagnet (Bisnovatyi-Kogan \& Komberg 1975), and models that 
call for a crustal nature of the 
magnetic field.

In this paper we modify the population synthesis calculation of 
paper I incorporating a
detailed physical model for the evolution of a crustal magnetic field
(using two boundary conditions to mimic expulsion or assimilation of the
field by the SC core), adding further effects that might alter the statistical
outcome. We also improve accuracy including the relativistic corrections 
(Burderi {\it et al.} 1999) necessary  to describe the spinup process when the
NS accretes matter from a disk, whose inner rim approaches the radius of the 
last quasi stable orbit. 
  
The  evolution in the $\mu$-$P$ plane is followed within a scenario of
recycling, constructed using our knowledge of accretion in low mass X-ray 
binaries (LMXBs). The study  is not aimed at correlating, on an evolutionary 
level, the observed population of LMXBs with the observed population of MSPs. 
MSPs may not be indeed the direct descendant of LMXBs: establishing this 
potential link or the link with LMBs forming through alternative evolutionary
channels (Kalogera \& Webbink 1996, 1998) would require the complete 
statistical knowledge  of the orbital parameters, of the mass ratios, of the
hydrodynamical processes guiding mass transfer and ultimately of 
the internal evolution and structure of the donor star
(see e.g. Ergma, Sarna \& Antipova 1998; Muslimov \& Sarna 1993). These are
aspects of 
a complex statistical calculation that is beyond our purposes.
We wish here to determine whether a minimal but possibly physically
consistent model for the recycling can provide indications
on the period distribution of NSs in the millisecond and sub-millisecond
domain, as guide line for future searches. Our aim is 
also to address a number of questions that we are outlining below.

In our population synthesis model we account for the evolution during the 
early phases when the NS in the LMB behaves as if isolated and later when 
fed by wind accretion. The mass transfer during the Roche Lobe Overflow (RLO)
phase is modelled just considering a range of accretion rates which is close to 
the one observed in LMXBs (Frank, King, \& Raine 1992; Webbink, Rappaport \&
Savonije 1983). In a number of models,
the accretion rate is treated as a constant over the RLO
phase: This is an approximation 
as its  actual behaviour 
follows a complex evolutionary pattern depending on the orbital parameters
and on the degree of mass and orbital angular momentum losses
(Ergma et al. 1998; Ergma \& Sarna 1996).
 
The increasing evidence that NSs in LMXBs may suffer phases 
of transient accretion due to thermal viscous instabilities in irradiation 
dominated disk (e.g. van Paradijs 1996; King, Kolb \& Szuszkiewicz 1997;
King, Frank, Kolb \& Ritter 1997) 
is suggestive that mass transfer onto a NS may not stop suddenly, at the end 
of the RLO: the star probably undergoes a progressive reduction of the mean 
accretion rate, modulated by alternate phases of higher and lower accretion.
The effects of this transient behaviour on the evolution of the magnetic field 
is not known yet (Page, Colpi, Geppert \& Possenti 1999,{\it in preparation}),
but, if the magnetic field is not too weak, any drop in  the accretion
rate implies the growth of the typical dimension of the magnetosphere 
(Bhattacharya \& van den Heuvel 1991, see $\S~2$ for details). Dragged by the 
rapidly spinning NS, the magnetosphere can transform in a centrifugal barrier, 
accelerating the matter at the inner rim of the accretion disk on 
superkeplerian orbits (propeller effect: Illarionov \& Sunyaev 1975); 
the extraction of angular momentum from the compact object that follows 
spins down the NS. {\it Does unsteady accretion at the end of the RLO phase 
vanish the effect of the previous phases of spinup? Is the potential population
of very rapidly spinning NSs reflected back to $P>P_{min}?$} 
Many physical ingredients necessary to fully describe the propeller induced 
spindown are poorly known or difficult to assess (e.g. the exact law for
the decrease of the mass transfer rate or the efficiency in the
extraction of the angular momentum from the NS to the propelled matter). 
So, we studied  the problem on a statistical basis,
parametrizing the most uncertain quantities.

The fastest rotating NSs detected so far have been discovered at radio 
frequencies, and statistical analyses based on current samples of MSPs 
(Cordes \& Chernoff 1997) have revealed that the pulsar distribution
is increasing toward short periods with best-fit minimum period slightly below
$P_{min}.$   Radio Searches for very rapidly 
spinning objects are now in progress.  In particular the large scale 
survey 
at the Northern Cross Radiotelescope near Medicina (D'Amico {\it el al.} 1998),
has a sensitivity profile relatively flat in the submillisecond period
range.  If successful, these searches can potentially
provide information on the EoS for nuclear matter (Burderi \& D'Amico
1997; Phinney \& Kulkarni 1994; Stergioulas \& Friedman 1995). If NSs can be 
recycled up to $P_{sh},$ the observed population can retain features 
of the EoS. Yet, many factors influence the recycling process,
(magnetic moment evolution, total amount of mass disposable for the accretion,
evolution of the mass transfer rate, propeller induced spin down and so on)
likely spreading the observed distribution to slower spin rates and perhaps, 
in extreme cases, almost  completely masking the effects of the EoS.
Thus, {\it can we determine, on a statistical basis, whether recycled rapidly 
rotating pulsars preserve signatures, in their $\mu$-$P$ diagram, for a
distinction between the EoSs?} 

Both types of evolution for a magnetic field of crustal origin allow, to a 
different degree, for the decay of $\mu$ at values lower than those typical 
of the observed MSPs sample, and probably below the threshold for
radio emission (Konar \& Bhattacharya 1998; Geppert \& Konenkov 1998; 
Sturrock 1971; Ruderman \& Sutherland 1975). In view of these searches 
an additional question is worth exploring: {\it Can we find a distinct 
feature in the distribution of NSs reflecting the boundary conditions at the 
crust-core interface? How significant is the production of weakly magnetized 
NSs during recycling?} As with the previous case, the statistical approach 
will help in clarifying whether the behaviour of highly condensed matter can 
be inferred from a close analysis of the NS period distribution.

In the $\S~2$ we outline the magneto-rotational evolution of a NS as
predicted by the recycling scenario, introducing  the parameters
adopted for the calculation. In the $\S~3$ we focus on the
evolution of the magnetic field residing in the crust of an accreting NS,
and describe the effects of different physical assumptions. In the $\S~4$
we present the results of the population synthesis calculations,
addressing one by one the questions risen in this $\S~1$. 
$\S~5$ contains the conclusions.

\section{\bf Spin Evolution Scenario}

\subsection{\it Physical model}

According to the recycling scenario (e.g. Lipunov 1992), the NSs in low mass 
binaries (LMBs) may experience the phase of  ejector, accretor, or 
propleller.
In the ejector phase, the NS  spins down only via radiation torque,
and we assume dipole emission according to the simple law
$$\dot P = 3.15\times 10^{-16}~~{\frac{\mu_{26}^2}{P}} {\rm {s~yr^{-1}}}
\eqno(1)$$
where $P$ is the neutron star rotation period in seconds, $\mu_{26}$ is the 
magnetic moment in units of $10^{26}$ Gcm$^3$.

In the propeller and accretor phases, 
matter penetrates down to the NS magnetosphere whose radius 
(Bhattacharya \& van den Heuvel 1991) 
is  
$$r_{mag}=9.8 \times 10^5~\phi ~\mu_{26}^{4/7}~{\dot{m}}^{-2/7}M^{-1/7}
R_{6}^{-2/7}  
{\rm cm}\eqno(2)$$
where $\phi$ is  estimated following Burderi et al. (1998); in equation (2)  
$M$ is the NS mass in solar masses, $R_6$ the static NS equatorial 
radius in units of 10$^6$ cm and $\dot{m}$  the accretion rate in
units of Eddington, i.e. $1.5\times10^{-8}~R_6 {\rm M_{\odot}}~$yr$^{-1}$.
In the propeller state, the uniform  angular velocity of the NS is higher than
the 
keplerian velocity at $r_{mag}$, and the magnetosphere acts  
as a centrifugal barrier (Illarionov \$ Sunyaev 1975). 
The infalling plasma is forced into super 
keplerian rotation and is propelled away: The NS loses
angular momentum and spins down.
The accretion  phase sets in everytime the magnetosphere does not act as a
barrier to matter approaching $r_{mag}$. 
In the statistical analysis of the RLO phase, we assume that the NS is fed
through a disk whose inner 
radius $r_{in}$ depends on the magnetic and rotational parameters 
of the NS. For high enough $\mu,$ the magnetic coupling between the disk and 
the star determines the extent of the angular momentum transfer: the
corresponding value of $r_{in}$ is $r_{mag}$.
In general $r_{mag}$ must be compared with 
both $R_{\Omega}$ and $r_{ms}$, where $R_{\Omega}$ is the physical equatorial
radius of the NS (accounting for the inflation due to very fast rotation)
and $r_{ms}$ is the radius of the last stable orbit. 
If the magnetic field is very low ($\mu \lsim 10^{26.5}$Gcm$^3$),
$r_{mag}$ is smaller than $R_{\Omega}$ (or $r_{ms}$) and the accretion
disk is truncated directly at the NS surface ($r_{in} = R_{\Omega}$) 
or at the last stable orbit ($r_{in} = r_{ms}$). 

The angular momentum balance relation used to follow the period evolution
of the NS is 
$${d(I_{\Omega} \Omega)\over dt} = g ~\dot{m}~ l_{in},\eqno(3)$$
where $I_{\Omega}$ is the moment of inertia of the rotating star,
$\Omega$ its angular velocity and $l_{in}$ the specific angular momentum 
of the accreting matter at the relevant inner radius $r_{in}$ of the disk.
The torque function $g=g(\Omega)$ accounts 
for the details of the angular momentum transfer 
between the NS magnetosphere and the accretion disk.
When $r_{in}=r_{mag}$ and $g=0.0$ the NS is on the so--called ``spinup line'', 
where it can accrete mass without modifying its angular momentum 
$I_{\Omega} \Omega$. 
If  the magnetospheric radius is inside  the marginally
stable radius 
 $r_{ms}$ or inside the star's radius, we set $r_{in}=r_{ms}$ (or
$=R_{\Omega}$), and 
$g$ is  equal to $1$.
When the NS is in a  propeller state 
the function  $g$ becomes negative (see \S2.2 for the values
adopted). 
If the position of the NS in the $\mu-P$ plane at the end of mass transfer 
is above the ``death line'' (Sturrock 1971, Ruderman \& Sutherland 1975),  
the NS shines as a {\it pulsar} and suffers secular
spin down
by magnetic dipole torques (eq. [1]). In this terminal phase,
the pulsar migrates to the right of the
$\mu-P$ diagram, drifting towards longer periods.

\subsection{\it Values adopted for the synthesized populations}

The NSs evolved in our population synthesis model have an initial
gravitational mass of $1.4~{\rm M_{\odot}}$. Cook, Shapiro \& Teukolsky (1994)
showed that all viable equations of state (EoSs) for nuclear matter allow 
for recycling a $1.4~{\rm M_{\odot}}$ unmagnetized NS to ultra-rapid spinning 
rates. $P \lsim 1.5~$ms are attained (before the mass-shedding instability 
disrupts the star) even in those models having a static maximum mass close 
to $1.4~{\rm M_{\odot}}$. However, the minimum attainable period depends 
on the specific equation of state.
We here investigated two EoSs: the FP-EoS (van Riper 1988) and the PS-EoS 
(Pandharipande \& Smith 1975), similar to the A-EoS and the L-EoS 
of paper I (see Arnett \& Bowers 1977 for the labelling of the EoSs). 
The FP is a soft equation of state 
(radius $R=1.06 \times 10^6$cm for a $1.4~{\rm M_{\odot}}$--static 
neutron star and central density $\rho_c=1.27\times 10^{15}$ g
cm$^{-3}$),
whilst the PS is one of the stiffer ($R=1.64 \times 10^6$cm 
and $\rho_c=3.6\times 10^{14}$ g cm$^{-3}$ for
the same canonical mass). 
In the most favourable circumstances and accreting $0.5~{\rm M_{\odot}}$ 
(an acceptable
amount for the baryonic mass transferred in a LMB), a NS can spin up 
to a period of $P_{sh}=0.73~$ms (for FP-EoS) 
and of $P_{sh}=1.40~$msec (for PS-EoS).  
These values have been computed according to Burderi {\it et al.} (1999),
who developed a semi--analytical model for studying the evolution of
the rotational period of a magnetic neutron star as a function of the 
accreted baryonic mass.
The  model permits the inclusion of general relativistic
effects and comprises: 
(i) the stellar deformation (circumferential radius) as a function of angular 
velocity, (ii) the variation of the moment of inertia 
in response to the mass load and to the increasing rotation,  
(iii) the location of the marginally stable orbit for 
the computation of the angular momentum transfer rate in low-field NSs; 
(iv) the decrease  of the mass shedding period for the compact
object, due to the increase of the  baryonic mass by  accretion.

In our statistical model of paper I, we evolved NSs in low mass
binaries (LMBs) through the five phases of the recycling scenario.
Here, we start with a population of NSs at the onset of the Roche Lobe
Overflow (RLO) phase. The adopted parameters are shown in Table~1.
Urpin, Geppert \& Konenkov (1998) calculated the tracks in the $\mu -
P$ plane for a sample of pulsars, assuming different hypotheses for the binary
evolution times, for the magnetic field 
decay and for the angular momentum transfer between the intervening 
plasma and the magnetosphere. Starting with pulsar parameters 
($\mu_{ini}$ and $P_{ini}$) in accordance with the most updated observational
analyses (Bhattacharya {\it et al.} 1992, Lorimer 1994, Hartman {\it et al.}
1997), they showed that, at the end of the wind phase, the typical NS magnetic
field is about $1.5-3$ orders of magnitude lower than at the onset of the 
pulsar phase (almost depending on the duration of wind accretion; see section
$\S 3$), whilst the period $P$ spreads between $\sim 0.1$ sec and $\sim 1000$
sec. During the RLO accretion the spinup line is reached after a time    
$\tau_{sp} \simeq 10^5 \div 10^6$ yr which is shorter than the duration of 
the RLO accretion phase. Thus, the rotational evolution of a NS is almost
independent of $P$ at the onset of the RLO phase.
Therefore, we adopted a simple flat distribution in period (see Table~1) as
input parameter for the calculations, verifying that the statistical 
outcome of the subsequent phases are  insensitive 
to the exact range of $P$ or/and to its detailed initial distribution.

The logarithm of the accretion rate $\dot m$ (in Eddington mass 
${\rm {\dot M}_E}$) during the RLO stationary phase is randomly selected 
from a gaussian distribution, with peak value ($0.1~{\rm {\dot M}_E}$) 
and spread $\sigma$ (half dex) compatible with the observed values.
In fact, the typical X-ray luminosities of the stationary LMXBs 
cluster in the interval $10^{36} \to 10^{38}$ erg/sec (van Paradijs 1995), 
implying accretion rate in the range 
$\sim 10^{-10} \to \sim 10^{-8} {\rm M_{\odot}/yr}$ (for masses 
$\sim 1.4~{\rm M_\odot}$). Uncertainties in distance of the sources and/or in
the radiative efficiency of the accretion process could affect these 
estimates. We note that the typical mean  accretion rates selected 
span the interval of values calculated by Ergma et al. (1998) 
within scenarios with conservative mass transfer. 
Low luminosity LMXBs may exist, yet undetected, 
so we explore lower mean rates on the synthesized population of recycled NSs.

As shown in Burderi {\it et al.} (1999), the minimum mass 
for spinning  a {\it magnetized} $\sim 1.40~{\rm M_\odot}$ NS  up to  
$P<10$ msec is $\sim 0.01~{\rm M_\odot}$,(this value doubles for unmagnetized
neutron stars). We adopt that value as lower limit to the total 
mass accreted, in  our synthesized populations. As regard to the upper limit, 
the maximum value of the mass transferred  
(i) must grant for the radial stability of the accreting NS 
and (ii) has to be compatible with the maximum duration 
of the RLO phase $\tau_{RLO}^{max}.$ For both EoSs, the first criterion 
is safely  satisfied by assuming an upper limit of 
$0.5~{\rm M_{\odot}}$ (as extrapolated from Cook, Shapiro \& Teukolsky 1994).
The second request introduces a free parameter.

We considered a flat probability distribution for the logarithm of
time of duration of the terminal phase of recycling, when the NS could emit 
as a radio. The minimum adopted value largely accounts for the time 
necessary to clean up the NS surroundings before the magnetic dipole emission
could set in again. 
The maximum time is limited by the age of the disk of 
the galaxy, for which we put an upper limit on the total duration of the 
entire recycling process. The value of $\tau^{max}_{RLO}$ could
be shortened if evaporation of the companion sets in (Muslimov \& Sarna 1993)
and we explore accordingly values as short as $\sim 10^7$ yr.
 
The behaviour of the function $g$ and the value at which it zeroes 
(dubbed as critical fastness parameter $\Omega_{s,crit}$) are still
uncertain. Therefore we adopted two different forms for $g$, following the 
suggestions of Ghosh \& Lamb (1991) and of Wang (1996 and reference therein).
However, both calculations considered only the 
positive branch of the $g$-function, which governs the transfer of angular
momentum from the disk to the NS. For a slowly rotating NS, their
estimates for the torque efficiency  are respectively
$g(\Omega=0)=1.17$ for Wang, and $g(\Omega=0)=1.40$ for Ghosh \& Lamb.
The predicted values of $\Omega_{s,crit}$ cover  a larger range 
due to the different approximations used in the calculation. 
However we noticed that our results do not change sensitively if  
$\Omega_{s,crit}$ varies in the interval $0.85 \div 1.00$ (in units
of the Keplerian angular velocity at the inner rim of the accretion disk)
compatible with the most updated estimates (but see Li \& Wang 1999
for a different view).

The shapes of the $g$-function near $\Omega_{s,crit}$ have to 
be assumed cautiously. No fully consistent calculation of the torque 
function is available yet for the description of
the propeller phase. In order to include this effect 
(especially at the end of the RLO phase), we extrapolated the known 
$g$ functions to their negative branch, introducing a central
symmetry with respect to their zero-point. This preserves continuity 
and derivability with the minimum number of further physical hypotheses.
Different semi-qualitative approximations (e.g. Menou, Esin \& Narayan
1998 adopted a linear relation between $g$ and $\Omega$) do not 
affect our statistical findings once $\Omega_{s,crit}$ is close to
$1.00$.

\subsection{\it Persistent and non-stationary accretion}

During the RLO stage, very extended phases of constant accretion rate 
are quite improbable. For example, Muslimov \& Sarna (1996) and 
Ergma et al. (1998) showed how $\dot m$ can vary in LMBs under different
hypothesis for the evolution of the system (conservative or non
conservative mass transfer, orbital angular momentum losses, initial
orbital period and/or secondary evloutionary stage). With the aim
of exploring the effect of a decreasing $\dot m$ on the population of
fastly spinning objects, we compared two possibilities: persistent accretion 
for a time $\tau_{RLO}$, or persistent accretion for a shorter time 
followed by a transient phase mimicking the quenching of the mass transfer.
We modelled the switching off of the accretion as a power law decay
for $\dot m$, investigating the effect of varying the two parameters:
(i) the ratio ${\cal F}_{que}$ between the duration of the quenching phase
with respect to the total RLO phase and (ii) the index $\Gamma_{que}$
of the power law. Adopted values for ${\cal F}_{que}$ range from $0.0 \to 0.5$
and for $\Gamma_{que}$ from $1$ ($\dot m \propto t^{-1}$) to $10$ 
(representative of an almost sudden switch off).
Probably the quenching of the accretion is not a smooth process, but
develops through alternate phases of higher and lower mass
transfer rate, transforming the NS in a transient source of X-rays.
Due to the short timescales of outburst and recurrence (with
respect to the duration of the whole quenching process), our
crude model could describe satisfactorily also this more realistic situation.
On the contrary, our model can not account for those cases 
in which 
the accretion rate initially decreases and later increases substantially
(in tight systems as indicated in Ergma et al. 1998;  for MSPs recycled in
Intermediate Mass Binaries
as recently suggested in Podsiadlowski \& Rappaport 1999).

We account for the angular momentum losses by propeller,
both for persistent and for non-stationary accretion during phase IV. 
An extended spin down phase is quite unlikely in the former case,
because a steady accretion rate and a reduced magnetic moment imply a very
small magnetosphere. On the contrary, if, at the end of the RLO phase, 
$\dot m$ decreases, the magnetospheric arm  can increase, allowing 
for an efficient transfer of angular momentum from the NS outside.

\section{\bf Magnetic field evolution in LMBs}  

\subsection{\it Physical model}

A basic assumption for the present investigation is the crustal origin of the 
neutron star magnetic field, i.e. it is created at the 
beginning of the NS's existence in its crustal layers. 
The onset of a thermomagnetic instability, which transforms heat into 
magnetic energy during the early phases, is an effective tool to produce 
strong fields in the crust of a NS (Urpin, Levshakov \& Yakovlev, 1986;
Wiebicke \& Geppert, 1996). Although that instability is not yet completely 
understood, it is a plausible mechanism which does not depend on special 
assumptions and may account for the observed variety in the NS
magnetic field  strengths.
Moreover, the assumption of a crustal magnetic field  is in
accordance with the general
accepted evolutionary scenarios for isolated NSs (Urpin \& Konenkov, 1997),
NSs in low mass binaries (Urpin, Geppert \& Konenkov, 1998), 
NSs in high mass binaries (Urpin, Konenkov \& Geppert, 1998) 
and millisecond pulsars (Geppert \& Konenkov, 1998).

Since, at the beginning of the RLO phase, the anisotropies of the 
conductivity $\sigma$ can be safely neglected, the evolution of the
MF 
is governed by the induction equation
$$\frac{\partial \vec{B}}{\partial t} = - \frac{c^{2}}{4 \pi}
\nabla \times \left( \frac{1}{\sigma} \nabla \times \vec{B}
\right) + \nabla \times ( \vec{v} \times \vec{B}) \;, 
\eqno(4)$$
where $\vec{v}$ is the flux velocity of the accreted matter 
through the crust.
When considering only the evolution of an axisymmetric 
dipolar poloidal field 
the induction equation can be simplified by introducing the vector potential 
$\vec{A} = (0, \,0, \,A_{\varphi})$ and choosing 
$A_{\varphi} = s(r,t) \sin \theta /r$. With the spherical radius $r$ and 
the polar angle $\theta$, both the radial and meridional field components can 
be expressed in terms of the quantity $s(r,t)$,
$$B_{r} = \frac{2 s}{r} \cos \theta \, , \,\,\, B_{\theta} = 
- \frac{\sin \theta}{r} \frac{\partial s}{\partial r}, \eqno(5)$$
which gives the maximum field strength at the NS surface ($r = R$) 
by its polar value for $B_{r}$, that is $B_{s}(t) = 2 s(R,t) / R$.
Although in the outer layers of the crust the inflow of matter channeled 
by the magnetic field is certainly not radial, the mass flow will
become nearly 
spherical symmetric in the higher density regions, where the 
currents are concentrated. In the
approximation of 
a purely radial mass flow, $\vec{v} = -v_{r} \vec{e_{r}}$, the induction 
equation can be transformed to 
$$\frac{\partial s}{\partial t} = \frac{c^{2}}{4 \pi \sigma} 
\left( \frac{\partial^{2} s}{\partial r^{2}} - \frac{2 s}{r^{2}} \right) 
- v_{r} \frac{\partial s}{\partial r} \,\,, \eqno(6)$$
where the flow velocity is given by the equation of continuity as 
$$v_{r} = \frac{\dot{M}}{4 \pi r^{2} \rho} \;. \eqno(7)$$
\noindent
and ${\dot M}={\dot m}{\rm {\dot M}_E}$.
While at the surface the standard boundary condition for a dipolar field 
applies, i.e. $R \partial s/\partial r + s = 0$, the inner boundary condition 
is more complicated and still a subject of scientific debates. In order to 
consider the possible different evolution of the NS magnetic field  
at the crust--core 
boundary we will apply two qualitatively different inner boundary conditions.
During the accretion process onto a NS its field  both diffuses and
is advected towards inner layers. It is generally accepted that the core 
of a NS 
older than $10^9$ yr is in a superfluid/superconductive (SF/SC) state, a state
which will not be changed even by the intense heating during the RLO phase. 
Thus, in case of $\vec{v} = 0$, at the crust--core boundary the correct 
boundary condition is $s = 0$, since the Meissner--Ochsenfeld effect prevents 
the magnetic field from diffusion into the core.
However, in case of $\vec{v} \ne 0$, the field, frozen in on
timescales of matter
flow, can cross the crust--core boundary and becomes assimilated in the core,
together with the former crustal material.
Our first boundary condition (BC I) assumes that, during 
the process of assimilation, $\vec{B}$  will be expelled immediately 
by the crustal material undergoing the SF/SC phase transition. 
Thus we define:
$${\rm BC~I~:}~~~~~~~~~s = 0~~~~~{\rm{at}}~~~~~r = r_{crust-core}~~. 
\eqno(8)$$
Alternatively we consider a second boundary condition (BC II), which allows an 
advection of the magnetic field into the core. There, it will be not
further decayed due to
SC but further advected as long as accretion lasts. This assumption results in:
$${\rm BC~II~:}~~~~~~~~s = 0~~~~~{\rm{at}}~~~~~r = 0~~. 
\eqno(9)$$
which may cause the occurrence of a residual field remaining constant
with time, 
({\it freezing} of the field) whereas BC I allows for an exponential
magnetic field decay 
down to zero. The situation assuming BC II has been described by 
Konar \& Bhattacharya (1997) and we apply also their trick to simulate 
an effectively infinite conductivity in the assimilated core region.

For the density profile we  assume that it will  not be changed by the 
accretion of matter onto the NS. That is certainly a good approximation 
as long as no more than $\sim 0.1~ {\rm M_{\odot}}$ has been accreted.
The chemical composition and density stratification can instead vary with
increasing mass load
($0.1 {\rm M_{\odot}} \rightarrow 0.5 {\rm M_{\odot}}$)
but these changes are considered unimportant, given the other approximations
introduced. 
 
At the beginning of the RLO, the currents maintaining the magnetic
field are already 
located close to the crust--core boundary 
($\rho_{crust-core}=2\times 10^{14}$ gcm$^{-3}$). 
Since even in the case of Eddington accretion the crust will not be heated up
above $5\times 10^8$ K, the crustal matter at densities larger than 
$10^{10}$ gcm$^{-3}$ is crystallized. Therefore, the relevant contributions
to the electric conductivity in the crust are due to 
electron--phonon and electron--impurity scattering. Electron--phonon 
interactions dominates the transport at high temperatures and relatively low 
densities, whereas the impurity concentration determines the conductivity at 
lower temperatures and larger densities. The accretion induced  
heating of 
the crust diminishes the electron--phonon conductivity considerably since 
it is $\propto T^{-1}$ above the Debye temperature and $\propto T^{-2}$ 
for lower $T$. Fujimoto {\it et al.} (1984), Miralda--Escude {\it et al.} 
(1990) and Zdunik {\it et al.} (1992)
calculated the crustal temperature as a function of the accretion rate. 
For both the stiff and the soft NS model, we used the following
fitting formula, which gives the dependence of the crustal temperature $T$ 
on the accretion rate in agreement with the numerical results of the above 
authors (see Urpin \& Geppert, 1996):
$$\log T = 7.887 + 0.528 \left[ 1 - e^{-0.899(q + 11)} \right], \eqno(10)$$
where $q = \log (\dot{M}$/M$_{\odot}$ yr$^{-1}$). 
For $\log (\dot{M}$/M$_{\odot}$ yr$^{-1}) > -9.5$, Eq. [10] 
yields probably a somewhat too small crustal temperature. We use the numerical
data for the phonon conductivity $\sigma_{ph}$ obtained by Itoh et al. (1993) 
and a simple analytical expression for the impurity conductivity 
$\sigma_{imp}$ derived by Yakovlev \& Urpin (1980).
The total conductivity is given by
$$\frac{1}{\sigma_{tot}} = \frac{1}{\sigma_{ph}} + \frac{1}{\sigma_{imp}} \;.
\eqno(11)$$
The lowest accretion rate we will consider for the RLO phase is 
$\log (\dot{M}$/M$_{\odot}$ yr$^{-1}) = -11$, which corresponds to a crustal 
temperature of about $8\times 10^7$K. For such and higher temperatures, 
$\sigma_{ph}$ is the dominating contribution to the conductivity in the 
whole crust. Besides on the temperature, density and impurity concentration, 
the conductivity depends on the chemical composition of the crust too. 
During an extended RLO phase the whole mass of the crust can be replaced 
by accreted matter. Therefore, instead to assume that the crust is composed 
of cold catalyzed matter we apply the chemical composition established by the 
pycnonuclear reactions in the course of the accretion 
(see Haensel \& Zdunik, 1990). The electric conductivities for the
weakest and the strongest RLO accretion rates considered in this paper 
are shown in Fig.~1.

The density profile for the soft--EoS NS is given by van Riper 1988 
(see Fig.~1 in Geppert \& Urpin 1994), whereas the density profile for 
the stiff--EoS NS is taken from Pandharipande \& Smith (1975). 
The EoS valid in the core determines the thickness of the 
crust and hence the scale of the crustal field. Hence, apart from the 
accretion driven decay mechanisms, it governs strongly the field evolution: 
the softer the EoS the faster the crustal field decays. This correlation is 
partly counterbalanced by relativistic effects which become stronger 
the softer the EoS is (Sengupta 1998).

In Figure~2 we represent (for $5\times 10^6 \leq \tau_{RLO} \leq 5\times 
10^8$ yr) the decay of the surface magnetic field  under the influence
of different accretion rates during the RLO phase both for soft and stiff EoS
and for the two boundary conditions considered here.
The BC I allows for an exponential decay  down to zero if the 
accretion period is extended enough . It is clearly seen that the larger 
the accretion rate the hotter the crust (see eq. [10]) and, hence, 
the more rapid the crustal field decays. The more rapid decay in case of the 
soft EoS in comparison with the stiff one is a result of the smaller spatial 
scale of the magnetic field. For both EoS we 
see that the assumption of BC II results in a much stronger surface field. 
Only for nearly Eddington accretion rates and soft EoS (lower right panel:
curve labelled by $-8.5$), the very rapid field decay is practically 
unaffected by the choice of the boundary condition. In that case, perhaps, 
effects of submergence play an important role because 
for such large accretion rates the characteristic time for the mass flow in 
the lower density layers of the crust is smaller than the ohmic 
(re--)diffusion time in that region.  For the soft EoS a trend in 
establishing a residual field is at most suggested for 
$\log (\dot{M}$/M$_{\odot}$ yr$^{-1}) = -10$ and $-9.5$.
The existence of residual fields seems to be much more evident for stiff EoS.
For the lowest accretion rates the velocity of mass
flow at the crust--core boundary is certainly too small to create residual
fields during $\tau_{RLO} \leq 5\times 10^8$yr and for the near Eddington 
accretion the effect of submergence acts. However, for 
$\log (\dot{M}$/M$_{\odot}$ yr$^{-1}) = -9, -9.5$ and $-10$, the development 
of a residual {\it frozen} field is clearly seen and its strength is 
positively correlated to the accretion rate, 
as Konar \& Bhattacharya (1998) recently pointed out.

\subsection{\it Values adopted for the synthesized populations}

The typical value of the magnetic moment  $\mu$ for the observed 
pulsar population lies in the range of $10^{29.5} \div 10^{31.5}$ Gcm$^3$ 
(Bhattacharya {\it et al.} 1992, Lorimer 1994, Hartman {\it et al.} 1997). 
During the subsequent evolutionary stages in a LMB, $\mu$ decays either by 
ohmic diffusion during the cooling of the NS (ejector and/or propeller phase) 
or by accretion driven decay mechanisms during the comparatively 
extended wind accretion phase, which can last up to $10^{10}$ years.
In these phases the depth of the crust initially penetrated by the magnetic 
field determines its strength. When the standard NS enters the RLO phase   
it has an age $\gsim 10^9$ yrs, and the field have had enough time to diffuse
down to the crust--core boundary, where the steep increase of the electric 
conductivity prevents  further inward diffusion, wherever the field was 
initially contained. Thus, at the onset of RLO, the field is located 
at the core-crust boundary, and its absolute strength is the parameter to be
varied in the population synthesis. Examining the various paths in the 
$\mu - P$ plane calculated either
by Geppert \& Konenkov (1998), or by Urpin, Geppert \& Konenkov (1998),
it appears that, at the end of the wind phase, 
the bulk of the evolved NSs group on the $\mu$-axis at values around 
$10^{28} \div 10^{29.5}$ Gcm$^3$. Such a clustering of the NSs, in a 
relatively narrow interval in $\mu,$ allows us to bar the uncertainties 
in the magnetic evolution of the objects during the first three stages of 
the recycling scenario. The values $\mu_0$ for the magnetic moment at the 
onset of the RLO era ($t_0^{\rm{RLO}}$) are selected from a gaussian 
distribution (see Table~1) with a mean chosen in that interval.

\section{\bf Results \& Discussion}

To execute the evolution we built a Monte Carlo code, typically
using 3,000 particles. The statistical analysis is performed only on those 
NSs reaching the so-called ``millisecond strip'' at the end of the
fifth stage ({\it radio}) of the recycling scenario, the ones  having 
period $P\leq 10.0$ ms and whichever value of the magnetic moment $\mu$. 
In accordance with the values of $P_{min}$ and $\mu_{min}$
(the weakest magnetic moment observed in PSRJ2317+1439; 
$\mu_{min}=7.3\times 10^{25}$ G cm$^3$), we divided our particles in four 
groups. Those filling the first quadrant in the millisecond strip 
($P\geq P_{min}$ and $\mu\geq\mu_{min}$)  behave as the known MSPs. 
Also the objects belonging to the second 
quadrant  ($P<P_{min}$ and $\mu\geq\mu_{min}$) should shines as pulsars
(see Burderi \& D'Amico 1997 
and the discussion of $\S~1$). 
The effective observability of the objects in the third quadrant 
($P<P_{min}$ and $\mu < \mu_{min}$) as radio sources represents 
instead a challenge  for the modern pulsar surveys. Most of them will   
be above the Chen and Ruderman (1993) ``death-line'', and might have
a bolometric luminosity comparable to that of the known MSPs.  
Thereafter we shortly refer as {\it sub-}MSPs to all the objects 
having $P<P_{min}$ and $\mu$ above the ``death-line''.  
Objects in the fourth quadrant ($P\geq P_{min}$ and $\mu < \mu_{min}$)
are probably radio quiet neutron stars (RQNSs), 
because they tend to be closer
to the theoretical ``death-line'', and they are in a period range which
was already searched with good sensitivity by the radio surveys.

\subsection{\it Are submillisecond pulsars a natural outcome of 
evolution in LMBs?}

In this paragraph, we discuss the results of our population synthesis
calculations  with no inclusion of the propeller effect.
We will refer thereafter as {\it standard} models.
In Fig.~3, we report the observed sample of
MSPs and the calculated populations for the two EoSs and 
the two adopted boundary conditions at the crust-core interface. The input
parameters are those of Table~1 with $\tau_{RLO}^{max}=5\times10^8$ yr. It
appears that objects with periods $P<P_{min}$ are present in a 
statistically significant number. In effect, though the detailed 
distributions of particles in the millisecond strip are influenced either 
by the EoS and by the magnetic field decay (see also Figs.~4 \& 5), 
{\it a tail of potential sub-}MSPs {\it always
emerges}.

In the synthetic sample for the {\it soft}-EoS, the
``barrier'' at $P_{sh}\simeq 0.7$ ms is clearly visible in Fig.~3, 
indicating that the field decay in longlived persistent LMBs proceeds fast 
enough to allow a significant fraction of objects to attain $P_{sh}$. 
The results are more clearly illustrated in Fig.~4
(solid line):  the {\it soft-EoS gives rise to period-distributions that 
increase rather steeply toward values smaller than 2} ms, irrespective to the 
adopted BCs. Instead, the boundary condition affects the distribution on $\mu$:
BC II produces a smaller number of objects with
low field, as the field initially decays but, when the
currents are advected toward the crust-core boundary, their decay is
halted and the field reaches a bottom value. In particular,
as already found in paper I where a BC of type II was considered,
the fourth quadrant is 
Fig.~5 is underpopulated (see also Tab.~2. 

The very {\it stiff}-EoS permits periods $P$ longer than $1.4$ ms, but the 
minimum observed period $P_{min}$ is already close to the ``barrier'' of 
mass shedding. Moreover, the field decay establishes on time scales longer 
than for a soft EoS and, on average, the sample of objects has a higher 
mean field. As a consequence, 
only fewer objects appear on the left of $P_{min}$ (Fig.~3).
NSs  with
$\mu<\mu_{min}$ are rare objects for a BC of type II, and 
the ratios of {\it sub}-MSPs to  MSPs are reduced with
respect to the soft EoS (for the boundary condition BC II we have 0.003 
for the stiff EoS and 0.573 for the soft one; see Table~2 for a complete 
review of the results).

Major differences appear also examining Fig.~4 (solid line): The period 
distribution for the stiff-EoS is much flatter than that for
the soft-EoS, displaying a broad maximum at $P\sim 3$ ms.
It was recently claimed that X-ray sources in LMBs (possible 
progenitors of the MSPs, see $\S~4.4$ for details) show rotational 
periods clustering in the interval $2 \to 4$ ms (White
\& Zhang 1997;  van der Klijs 1998). This effect could be explained 
introducing a fine tuned relation between $\mu$ and $\dot m$ 
($\mu \propto {\dot{m}}^{1/2}$: White \& Zhang 1997). 
Alternatively, gravitational waves emission has been invoked: 
either due to the growing of {\it r}-mode instability 
in the accreting NS (Andersson, Kokkotas \& Stergioulas 1998) or due
to a thermally induced quadrupole moment in the NS crust (Bildsten 1998). 
Here we note that such a clustering can be a natural statistical outcome 
of the recycling process.

\subsection{\it Is unsteady or transient
accretion in LMBs a threat to the existence of {\it sub-}MSPs?}
 
The suggestion that NSs in low mass binaries may become transiently accreting
sources after having experienced a phase of persistent accretion
prompted us to consider the effect of decreasing the mean accretion rate
$\dot m$ to mimic either the quenching of the accretion phase or
the phases of quiescence in transient sources.
If the magnetic field is high enough for the magnetosphere to affect the
flow, the propelled matter may transfer outward the angular momentum of 
the NS. Therefore, we have considered the possibility that the NSs spend 
a fraction of the RLO phase in transit to final quiescence.
The quenching of the accretion rate was modelled with a simple power law 
in time with index ${\Gamma}_{que}$ lasting a 
fraction ${\cal F}_{que}$ of the duration of the RLO phase. The 
Figures~4, 6, 7 \& 8 and the Table~2 show a summary of the results.

In Fig.~4 the dashed areas give the distributions of the NSs
at the end of evolution, including a strong propeller phase
(${\cal F}_{que}=0.50; \Gamma_{que}=8$) during the quenching of the RLO.
The emerging distributions are depleted of the fastest rotating NSs. 
However, it is remarkable that {\it for the soft EoS the 
distributions preserve a maximum just about $P_{min}.$}

In Fig.~6 we display the distribution on $P$ of our synthesized populations
at the end of the RLO phase (including strong propeller: dashed lines)
and at the end of the magnetic dipole emission phase (dashed areas).
It appears that dipole losses produce a clear shift 
of the entire population toward longer periods. However, it affects
preferentially the distribution of NSs at periods longer than $\sim$4 ms.
In our simulation we have noticed that dipole losses reduce considerably
the number of NSs in the millisecond strip (especially for $\mu>\mu_{min}$).
In particular Fig.~7 shows that dipole losses are much more effective than a
mild propeller (${\cal F}_{que}=0.25; \Gamma_{que}=1$)
in drifting the overall population at $P>10$ ms, whereas 
it only slightly influences the very rapidly spinning objects.

Figure~7 allows also for the comparison of the different number of objects 
synthesized in the millisecond strip with different models. 
These numbers mainly depend on the timescale for $\mu$ to decay 
down to $\sim 10^{27.5}$ Gcm$^3$, the value requested for a particle 
to enter the millisecond strip: the shorter the decay timescale is, the 
more filled is the millisecond strip. With BC I, the soft EoS gives a number 
of objects that is about 50\% greater than the stiff EoS. 
For BC II such a distinction is smoothed except for a very short RLO phase 
($\tau_{RLO}^{max}\simeq 5\times 10^7$ yr), for which the freezing of  $\mu$ 
is not operating yet.

Comparing Fig.~8 with Fig.~5, we note that {\it a strong propeller can
threaten the formation  of NSs  with $P<P_{min}$ and $\mu>\mu_{min}$}
especially in the case  of the stiff-EoS (see Tab.~2).
For objects with $\mu<\mu_{min}$, the spin-down torque is weaker:
accordingly, the III quadrant is less depleted than the second one and 
a non number of low magnetic field fastly spinning NSs can survive 
at periods $\simeq 1$ ms, for the soft EoS.

Our statistical analysis can provide also information on the NS mass 
distribution as a function of $P$ at the end of evolution (Fig.~9). 
We find that the mass function steepens toward high values  
when $P$ falls below $P_{min}$: the distribution approaches  
values close to $M\sim 1.7 \div 1.8 M_{\odot}$ as a large mass 
deposition is required to spin a NS to ultra short periods (we defer to 
Burderi {\it et al.} 1999 for a discussion on the minimum mass of NSs 
close to $P_{sh}$). LMBs could harbour NSs with such high values of the mass and
a first indication comes from Casares, Charles  \& Kuulkers (1998) who inferred
for Cyg X-2 a lower  mass limit of
$1.88~M_{\odot}.$ 
The action of the propeller during evolution has 
the clear effect of inhibiting the mass infall: the mass distribution 
is only slightly affected for the soft EoS, while for the stiff EoS 
the difference is more pronounced.

\subsection{\it A tool for discriminating 
EoS and  $\mu$-decay ?}

In this paragraph we summarize the results of  our statistical analysis 
considering in particular those objects having $P<P_{min}.$
Table~2 collects the main results derived for the different
scenarios (indicated in column ten).
{\it Standard} evolution is that discussed in $\S~4.1$. The other cases 
comprise a {\it propeller} phase at the end of the RLO ($\S~4.2$): here we 
adopted a propeller of intermediate effectiveness 
(${\cal F}_{que}=0.25$ and $\Gamma_{que}=1$). 
To account for the uncertainties in the distribution of the accretion
rates and/or in the maximum duration of the RLO stage, we run synthesis
calculations either in the case of a {\it low $\dot m$} (${\dot m}=0.01 
{\rm~{\dot M}_E}$) or in the case of a {\it short}  RLO phase 
($\tau_{RLO}^{max}=5 \times 10^7$ yr). Finally we combined the last two
hypotheses. 

The results are grouped by columns according to the adopted BC
and by rows as regard to the EoS. In particular, columns 2 (for BC I) and 
6 (for BC II) give the number of MSPs  produced during the runs,
normalized to the number of MSPs (set equal to 1000) synthesized 
in the case labelled by (iii) in the eleventh column. 
These figures  give an estimate of the efficiency of the recycling process 
in producing objects in the first quadrant. The third (for BC I) and the 
seventh columns (BC II) give the fraction of NSs 
in quadrant II relative to I. 
The fourth and the eighth columns (respectively for BC I and BC II) 
report 
the fraction (II+III)/I while in brackets we give  $f_{sub},$ i.e.,  the
fraction 
of NSs with $P<P_{min}$ and $\mu$
lying
above the death line relative to the sample of NSs filling quadrant I.

The estimate of $f_{sub}$ can give indication on the fraction of 
{\it sub-}MSPs relative to the observed MSPs: if selection criteria 
similar to those found in the interval  $P_{min}<P<10.0$ ms apply as well 
at periods 
below $P_{min}$, $f_{sub}$ provides an estimate of the number of pulsars 
that might be observable below $P_{min}.$ We find that $f_{\it sub}$ varies 
between 0\%$\to\sim$57\%; the lower figure applies only to the
case of an EoS of extreme stiffness: these objects should be observable 
with experiments like the survey in progress at the Northern Cross 
radio telescope near Medicina (D'Amico {\it et al.},1998), which 
has similar sensitivity in the MSPs and in the {\it sub}-MSPs range. 

Inspection of the Table~2 suggests that the EoS for the nuclear
matter has a major role in determining the ratio $f_{sub}$: the formation 
of a {\it sub}-MSP is quite a rare event, for the very stiff EoS. 
On the contrary, in the case of a moderately soft 
EoS, the {\it sub}-MSP population is significant, irrespective
to the details of the evolutionary scenario. 
Thus,  the detection of a significant number of {\it sub}-MSPs 
would be strongly suggestive of a not too stiff EoS. 
A clear distinction in the models can be seen comparing case (iv) with (ix).
If a moderate propeller takes place and the typical RLO timescale is shorter 
than usually assumed, for a soft-EoS $f_{sub}\simeq$30\% 
(irrespective the BC), whilst no {\it sub}-MSPs could appear if the nuclear 
matter behaves as predicted by the stiff-EoS.
The details of the evolution can blend the outcome of the
statistical analysis preventing a clear distinction between 
the equation of states when based on the simple ratio $f_{sub}$
(e.g. cases (i), (ii) and (iii) versus (x) for the BC I).
The boundary condition for the magnetic field at the crust-core interface
weakly affects the figures of Table~2 (except for case (viii)) when 
considering only the objects filling the I and the II quadrant 
(compare the columns 2 and 3 with the columns 6 and 7).
Due to their effects on the decay of the surface magnetic field,
{\it the differences between the two BCs are revealed mainly in the
synthesized populations having} $\mu<\mu_{min}.$ 

\subsection{\it Very low magnetic field NSs spinning at millisecond
periods?}

Except for the most unfavorable case (very stiff-EoS \& BC II), all our
synthesis runs produce a significant amount of objects with  $\mu<\mu_{min}$
(see III \& IV quadrants in the Figs.~5 \& 8). Their period distribution 
is very sensitive
to the adopted internal boundary condition for the magnetic field diffusion
at the crust--core interface.
BC II clearly favours  fastly rotating objects ($P\lsim 1.5$ ms), the 
IV quadrant in Figs.~5 \& 8 being (scarcely) populated only in the case of a 
very strong propeller. For BC I the distributions are broader and the 
stiffer the EoS is the more depleted is the III quadrant.
In principle, such a distinction could allow a discrimination among 
the two BCs.

On the observational point of view, very low magnetized NSs 
in the IV quadrant (upon which we focus in this paragraph)
might be elusive sources. Irrespective to the selection effects or survey
sensitivity 
thresholds, these objects might be rather close
to the theoretical ``death line'' to be observable as radio, 
and it is not clear if we can observe them in the X-ray band.
 
Recently a neutron star rotating in the millisecond interval (SAX J1808.4-3658
with $P=2.49$ ms) has been detected in X-rays (Wijnands \& van der Klijs 1998).
It is in a binary system with a low mass companion (Chakrabarty \& Morgan 1998)
which transfers mass to the compact object. As the rotational period is known, 
one can constrain the surface magnetic field of the accreting NS
(Psaltis \& Chakrabarty, 1998). An upper limit on $\mu$ relies on the 
requirement (a) that accretion is not centrifugally inhibited at the 
minimum accretion rate at which coherent pulsations are detected.
The standard disk-magnetosphere model (Pringle \& Rees 1972) interprets 
coherent X-ray pulsations as due to a 
complete or a partial funnelling of the disk plasma along the NS magnetic 
field lines. Many improvements had to be added to this model (Ghosh \& Lamb 
1991 for a review) to account for the variety of observations 
(e.g. Bildsten 1997); but in this framework, (b) a rotating NS can 
appear as an accretion powered pulsar only if the magnetic moment is strong 
enough to terminate the Keplerian disk above the stellar surface 
(Psaltis \& Chakrabarty 1998; Burderi \& King 1998). 
This requirement translates in a lower boundary on 
the possible value of $\mu.$ For the case of SAX J1808.4-3658, Psaltis 
\& Chakrabarty (1998) found $3\times 10^{25}\lsim\mu\lsim 10^{27}$
Gcm$^3$, with a preferred value about a few times $10^{26}$ Gcm$^3$.
So, this object seems to be placed near the separation of the quadrants
I and IV. Alternative models, not related or poorly related to $\mu$, could
explain coherent X-ray modulation at the NS spin frequency 
(e.g. azimuthal variation in optical depth for absorption or scattering) 
from which no further constraints on $\mu$ arise.

Already from the half of the nineties, the spin periods
for a handful of NSs in low mass X-ray binaries have been inferred  
either from coherent pulsations during Type-I bursts or, indirectly, 
from the so-called kHz-QPOs (van der Klijs 1998 for a review).
In the latter case, the paradigm is the so-called beat frequency model,
in which the frequency difference $\Delta \nu_{qpo}$ among the two peaks 
appearing in the X-ray power spectrum of the source is representative 
of the spin frequency of the NS or of an overtone (Miller, Lamb \& Psaltis
1998). From the inferred spin, and assuming that these NSs
rotate at about their equilibrium period (that is, near their competing spinup
line), one can estimate (through condition (a)) the value of $\mu$. 
As a representation, in Fig.~3, we display ({\it open squares}) the positions 
of the NSs of this group in the $\mu - P$ plane, 
adopting the values of White \& Zhang (1997 and reference therein for 
a discussion about the errors in the determination of $P$ and $\mu$).
It appears that no X-ray source shows $\mu<\mu_{min}$.

Prompted by the evidence of a not constant value for $\Delta \nu_{qpo}$ 
(Psaltis {\it et al.} 1998; Mendez \& van der Klis 1999), other models for 
explaining the kHz-QPOs have been proposed. For instance, 
Stella \& Vietri (1999) interpret the upper QPO 
frequency $\nu_2$ as due to matter inhomogeneities orbiting 
the NS at the inner disk boundary (just as in the standard beat frequency
model), whilst the lower QPO frequency 
$\nu_1$ is produced by the periastron precession at the inner edge 
of the accretion disk. Apart from negligible corrections (due to the effect
of the NS rotation on the exterior metric), the observed  
difference $\Delta \nu_{qpo}=\nu_2 - \nu_1$ is 
unrelated to the spin frequency of the NS. Being $P$ not constrained by
the observation,  condition (b) for a lower limit on $\mu$ 
does not apply anymore. Hence  we argue that also 
the magnetic moment estimates become questionable in this framework.
As a consequence, some of the sources shown in Fig.~3 could be misplaced 
in the $\mu - P$ plane and, when accretion will halt, they could 
eventually enter the category of low magnetic field NSs.
A similar suggestion was raised by Lai (1998), who
constructed refined slim disk models in LMBs, incorporating
the effects of both magnetic field and general relativity:
for the kHz-QPO sources he found values of the surface magnetic 
field smaller than the previous authors. If this is the case, 
the values of $\mu$ in kHz-QPO sources ($\lsim 10^{25}$ Gcm$^3$) 
are systematically weaker than those of MSPs sample and
these X-ray emitting objects would populate our IV quadrant.

As a general comments on the observability of low magnetic field NSs
in the X-ray band, we note that if the capability of coherently emitting 
X-rays pulses requires high enough values of $\mu$, missing rotating NSs 
with $P$ close to a millisecond  would not be surprising: 
as already shown in Burderi {\it et al.} (1999) and here confirmed, only
a rapid and substantial decay of $\mu$ allows a NS to reach $P<P_{min}$. 
Therefore, low magnetic field fastly spinning NSs could be just those not 
showing coherent X-ray pulsations. Instead, the signature of very rapidly 
spinning objects might hopefully emerge from low amplitude features in 
their power spectrum, related to not coherent modulation of the X-ray 
brightness. 

\section{\bf Conclusions}

\noindent
The population synthesis calculation has shown that:

\begin{itemize}
\item[1.] 
Detailed models for the decay of the crustal magnetic field,
including also refined relativistic corrections,
show the presence of a tail in the period distribution of
the synthetic NSs population at periods shorter than $1.558$ ms.

\item[2.]  
For the soft EoS, and irrespective to the boundary condition 
at the crust-core interface for the magnetic
field evolution, recycling in LMBs gives rise to a NS  distribution 
which is increasing towards shorter periods, and a clear
barrier is present at the minimum period for mass-shedding. 
For the stiff EoS, the distribution is flatter.

\item[3.]
If NSs at the end of the persistent accretion in a LMB experience a phase
of smooth decrease of the accretion rate 
(to mimic transient sources and/or quenching of accretion), the 
magnetospheric propeller produces a depletion of fastly spinning NSs
but, at least for the soft EoS, it preserves a distribution
that peaks at periods $\sim~{\rm 1.5~ms}$.

\item[4.] 
The estimated fraction of  {\it sub-}MSPs over the entire
MSPs population varies  between 0 to $\simeq 50$\%. The adopted EoS
and the parameters for the recycling 
play an important role in determining those percentages.
The detection 
of such rapid rotating compact objects represents a challenge for 
the modern searches.

\item[5.]
The models for the decay of a crustal magnetic field predict the existence
of spun up NSs with very low magnetic moment: their period distribution
is a neat signature for the physics at the crust--core interface.

\end{itemize}

\noindent
The first two authors want to acknowledge the 
hospitality of the members of the {\sl Astrophysikalisches Institut Potsdam} 
during the early phase of this work.

\newpage

\begin{figure}
\centering
\psfig{file=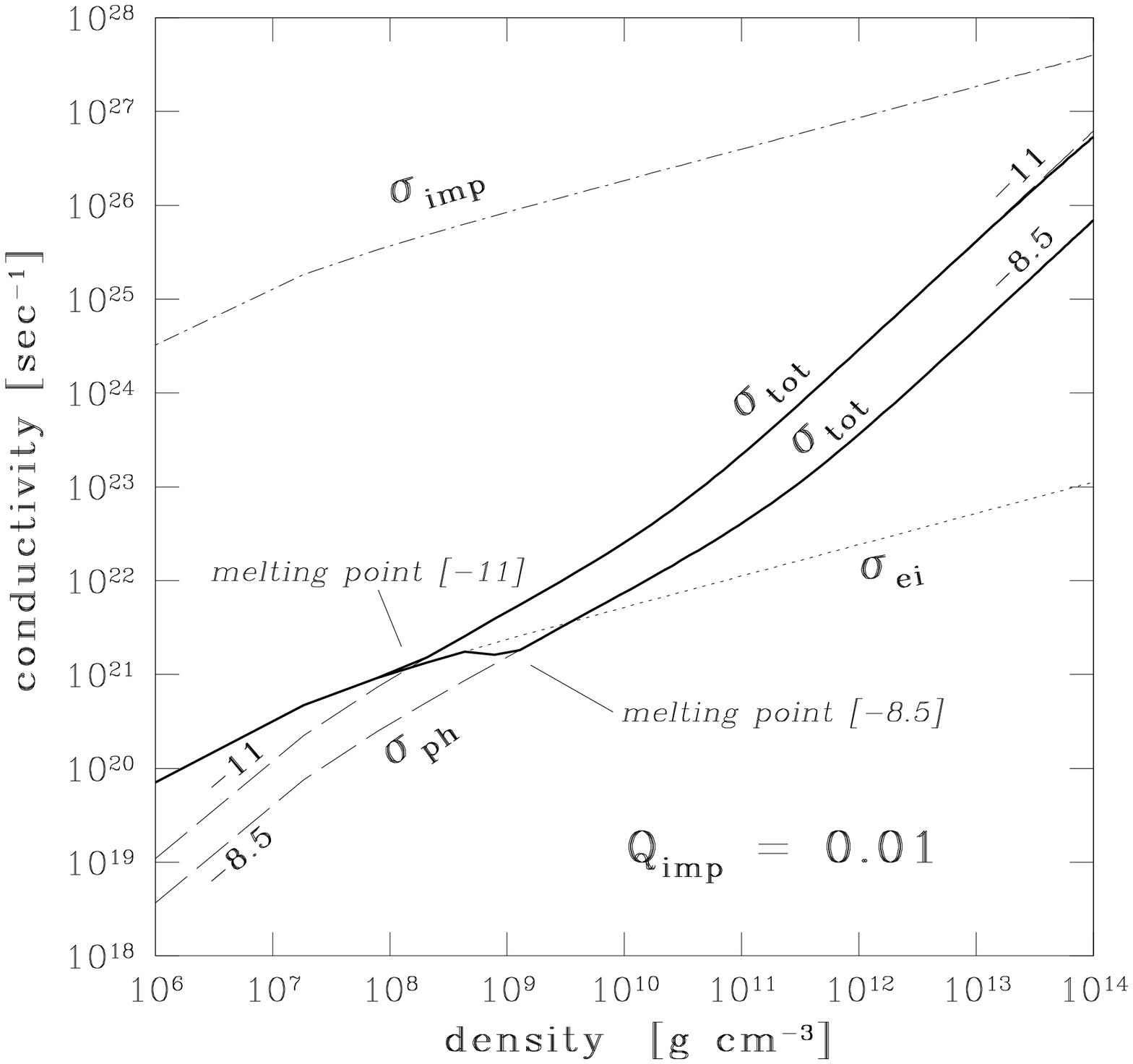,width=15cm}
\caption
{The electric conductivity in the crust as a function of the 
density for $\log (\dot{M}$/M$_{\odot}$ yr$^{-1}) = -8.5$ and $-11$. 
The {\it heavy solid lines} correspond to the total conductivity 
$\sigma_{tot}$. Since the impurity conductivity ($\sigma_{imp}$, 
{\it dot-dashed line}) is not dependent on the temperature, it is the same 
for both accretion rates. The conductivity in the liquid region ($\sigma_{ei}$,
{\it dotted line}) is determined by electron--ion collisions and also 
practically not dependent on $T$, contrary to the phonon conductivity
($\sigma_{ph}$, {\it dashed lines}), which dominates $\sigma_{tot}$ at
highest densities. The shift of the melting point towards higher densities 
for higher accretion rates is seen.
}
\end{figure}

\begin{figure}
\centering
\psfig{file=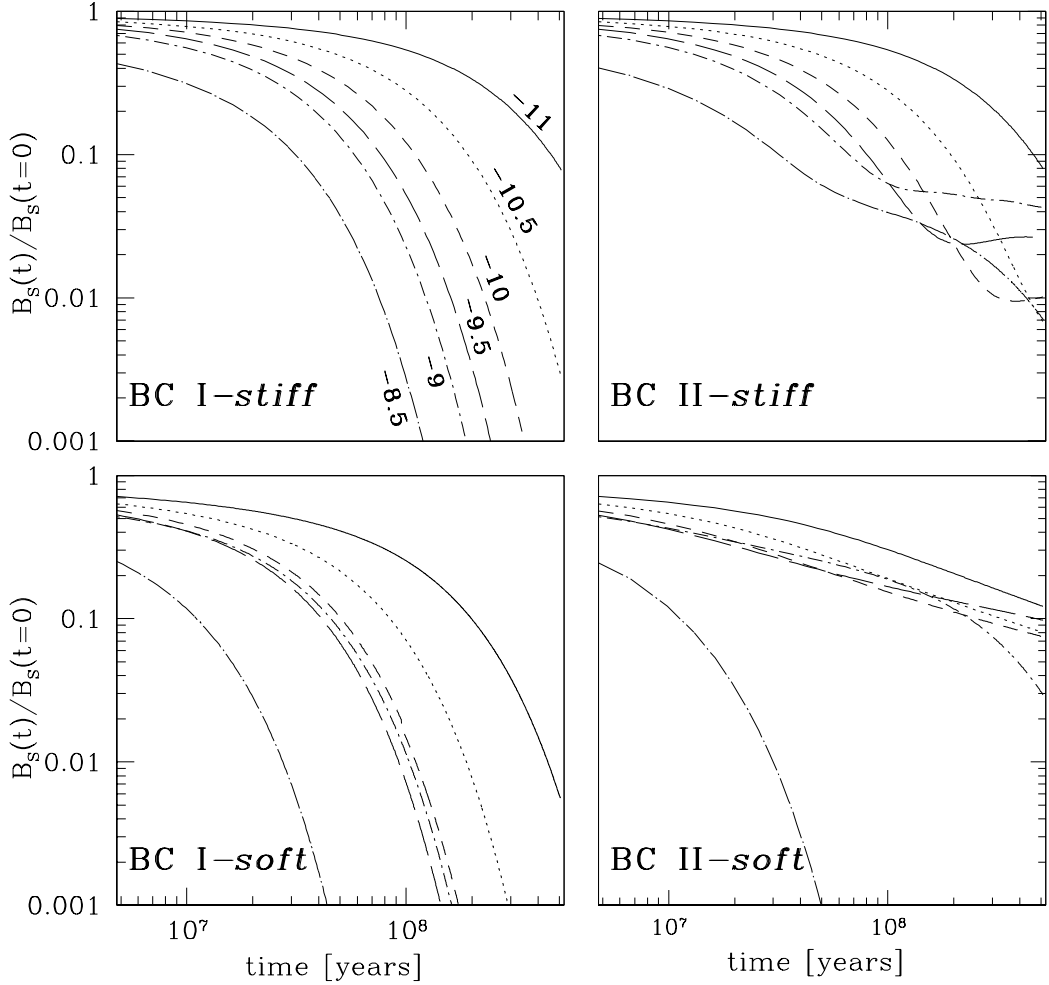,width=15cm}
\caption
{The dependence of the surface magnetic
field, $B_{s}(t)$, on the
time, normalized on the surface magnetic field strength at 
the beginning of the RLO phase ($t=0$). The different accretion rates 
are labelled according to $\log (\dot{M}$/M$_{\odot}$ yr$^{-1})$.
The four explored cases are located here as in all the following figures:
stiff-EoS on the upper panels, soft-EoS on the lower ones;
boundary condition BC I on the left panels and BC II on the right.
}
\end{figure}

\begin{figure}
\centering
\psfig{file=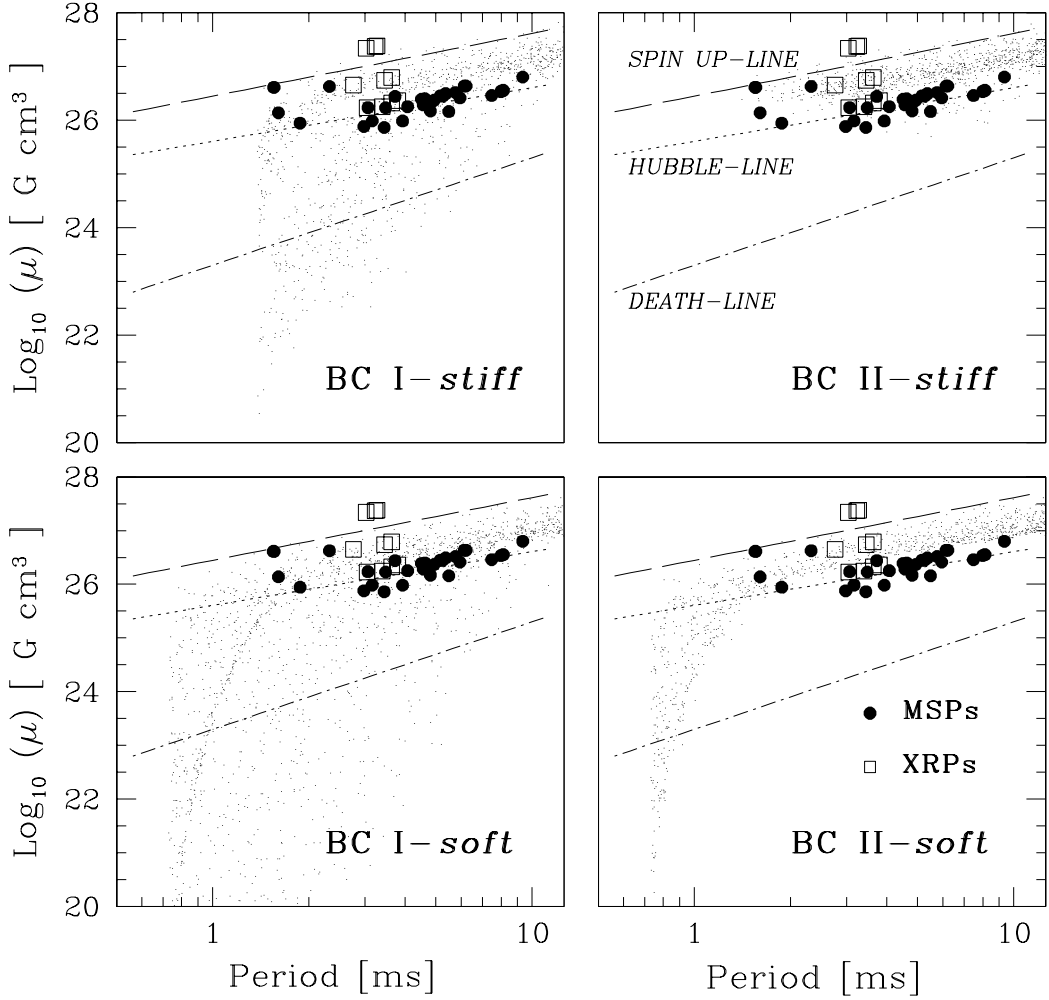,width=15cm}
\caption
{Statistical properties of the synthesized populations in the 
$\mu-P$ plane, for $<{\dot m}>=0.1$ and 
$\tau_{RLO}^{max}~=~5\times10^8~{\rm yr}$. {\it Full dots} represent the 
sample of detected MSPs (online catalogue), while {\it open squares} represent 
the NSs in LMBs for which the period and the magnetic field
have been inferred either from the kHzQPOs seen in their X-ray power
spectrum, or from coherent pulsations during burst, as reported in 
White \& Zhang (1997). 
}
\end{figure}

\begin{figure}
\centering
\psfig{file=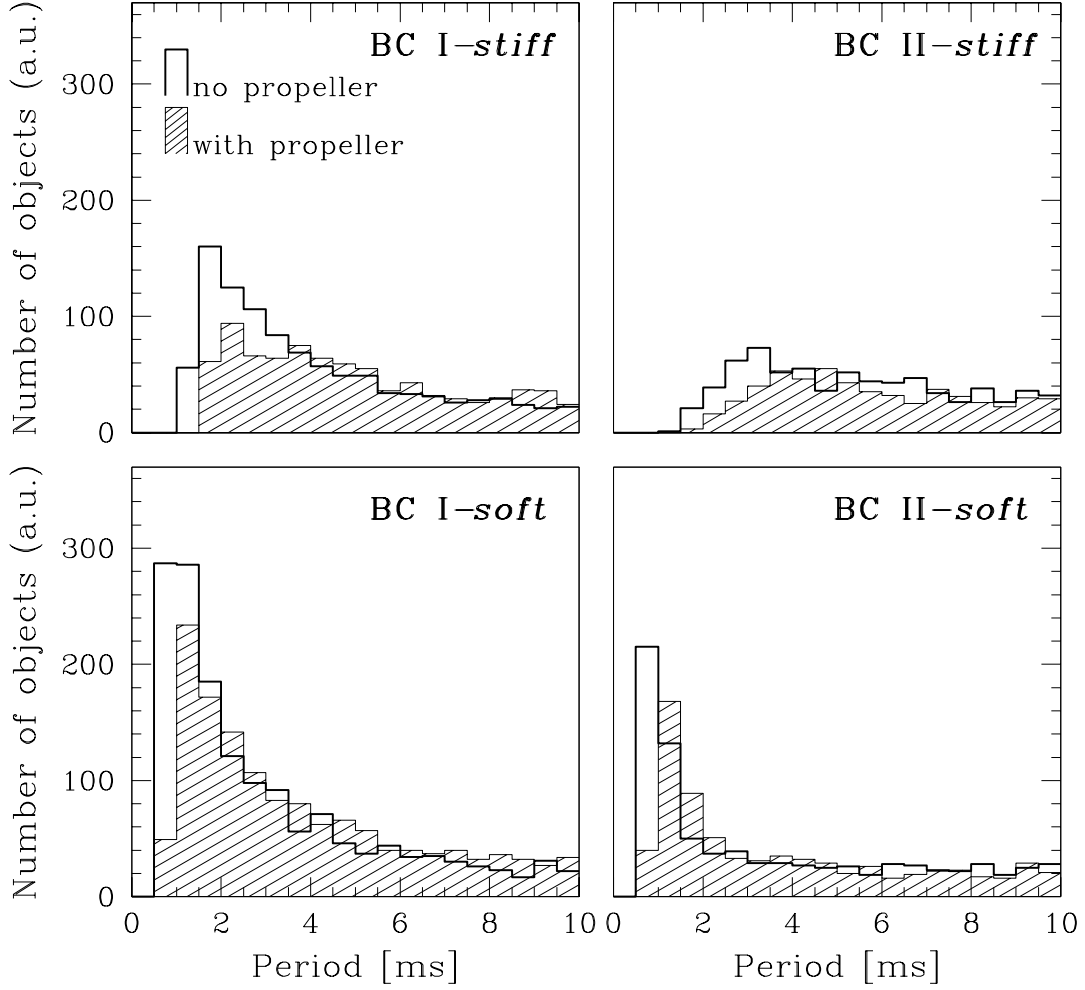,width=15cm}
\caption
{Calculated distribution of millisecond NSs 
as a function of the spin period $P.$ The selected interval is 
$P_{sh}<P<10$ ms and $\mu$ is let vary over the whole range. 
{\it Solid line} denotes the distribution in absence of propeller, whilst 
{\it dashed area} give the one with a strong propeller effect included 
$({\cal F}_{que}=0.50$ and $\Gamma_{que}=8).$ 
The absolute number of objects is in arbitrary units.
}
\end{figure}

\begin{figure}
\centering
\psfig{file=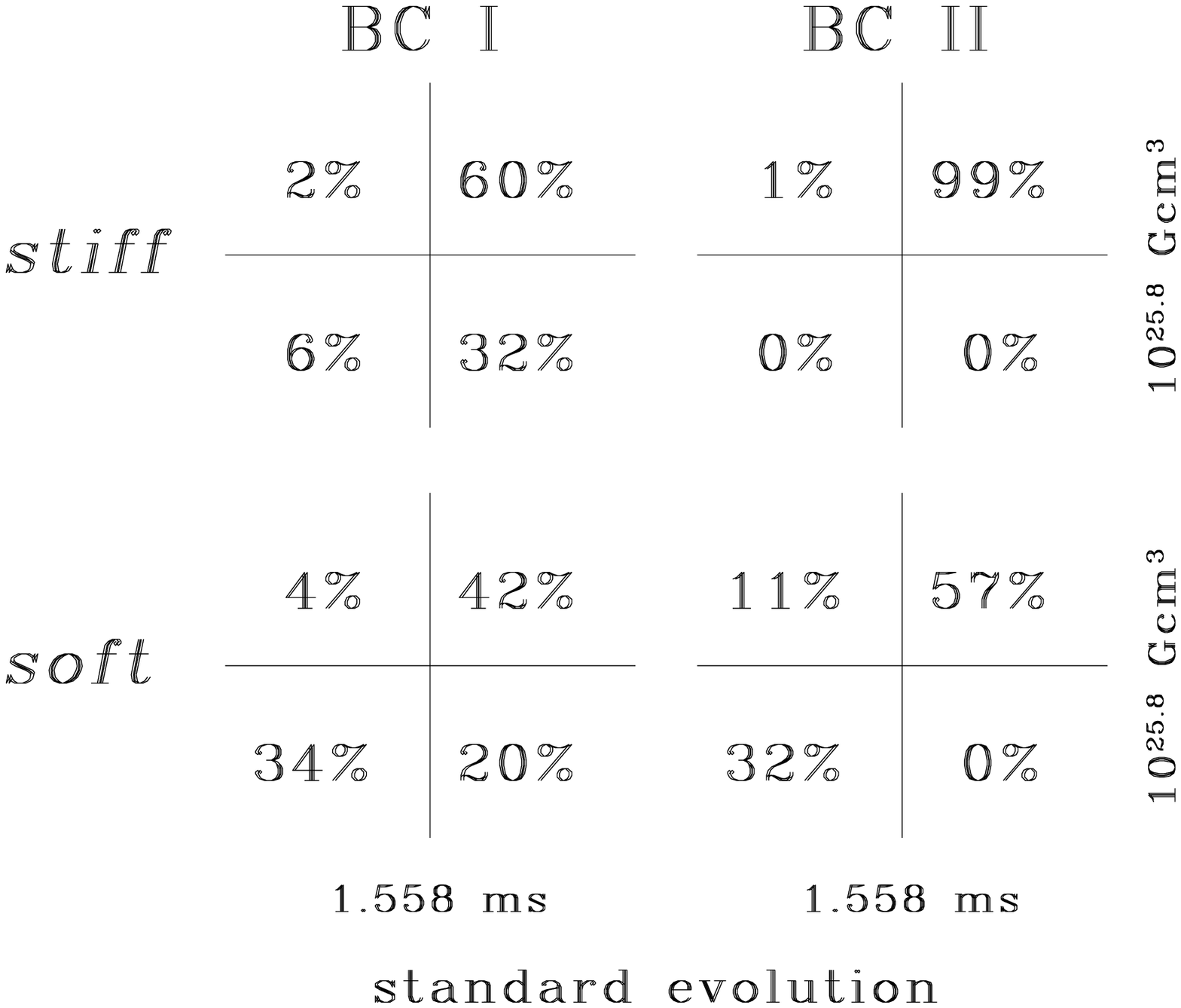,width=15cm}
\caption
{Distributions of the synthesized NSs, 
derived normalizing the sample to the total number of stars with 
$P<10~{\rm ms}$. We have divided the $\mu-P$ plane in four regions.
As a guideline the upper left number in each cross gives the percentage of
objects having $P<P_{min}$ and $\mu>\mu_{min}$ (the typical variance is
about 1\%). On the left side, we have indicated the EoS used. 
The crosses in the first column refer to the boundary condition BC I, 
whereas those in the second column to BC II.
}
\end{figure}

\begin{figure}
\centering
\psfig{file=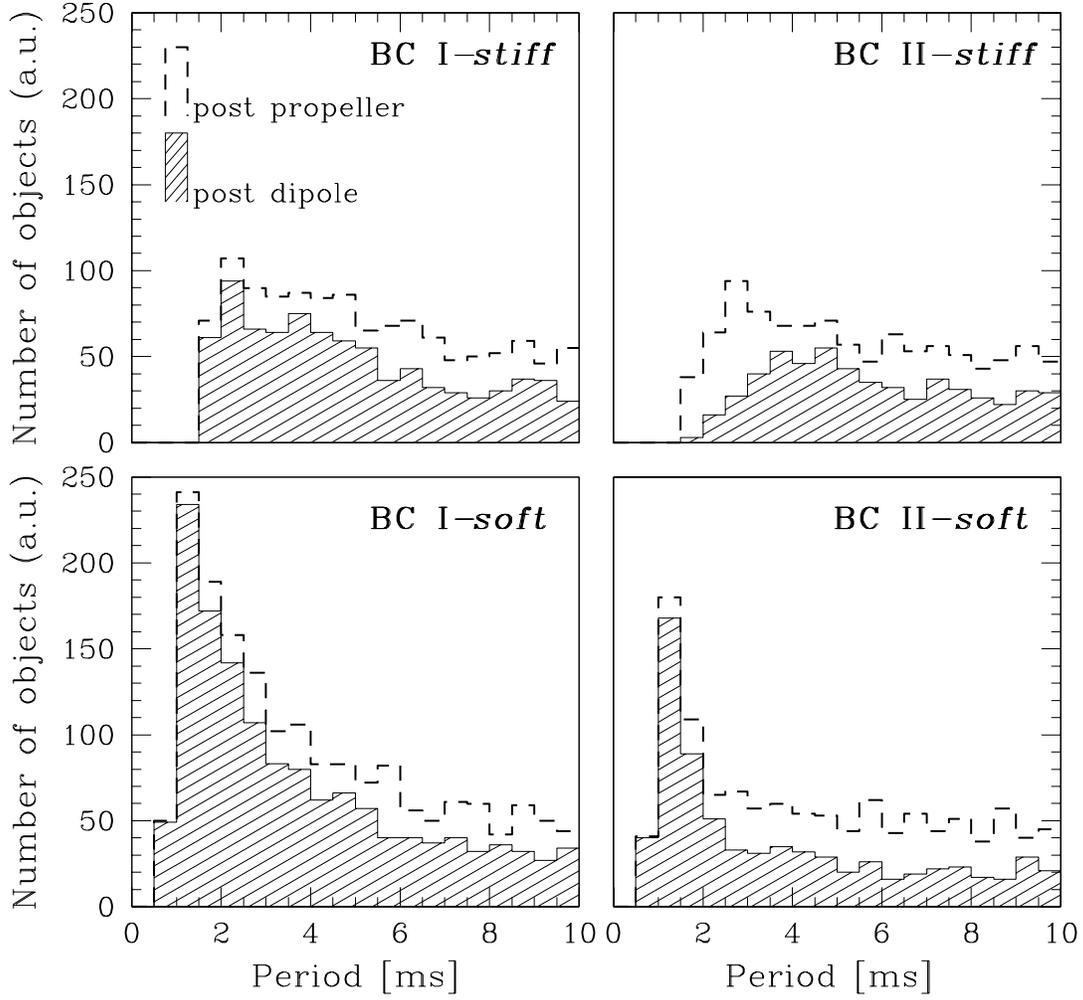,width=15cm}
\caption
{The four panels depict the NSs distributions at the end of the
propeller phase (dashed lines), and at the end of the magnetic dipole phase
(dashed area). $\mu$ is kept constant during the radio phase.
The propeller is mimicked as for the case of Fig.~4. 
}
\end{figure}

\begin{figure}
\centering
\psfig{file=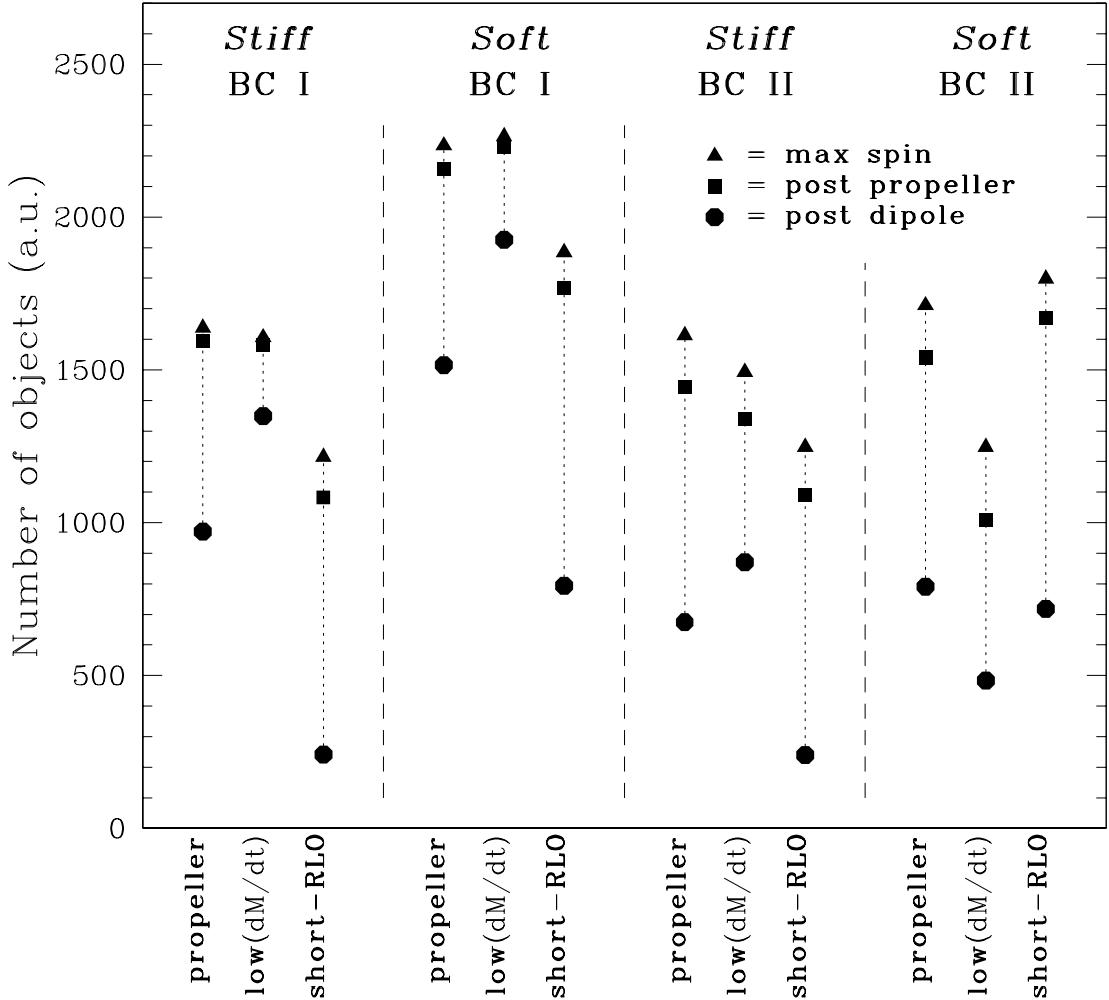,width=15cm}
\caption
{Number of synthesized objects with 
$P<10$ ms at different stages of their evolution and for different models 
as indicated in the labels. ``Propeller'' model refers to 
${\tau_{RLO}^{max}}=5\times~10^8$ yr, $<{\dot m}>=0.1$ in Eddington units,
${\cal F}_{que}=0.25$, $\Gamma_{que}=1$. 
For the ``Low(dM/dt)'' model we used $<{\dot m}>=0.01$;
while for the ``short-RLO''$~$ model we adopted  
${\tau_{RLO}^{max}}=5\times~10^7$ yr.
{\it Triangles} give the number of NSs at the end of the 
spinup phase during the RLO; {\it squares} give the number of NSs at the end 
of the propeller phase; {\it big dots} give the number of NSs at the end of 
the radio phase. The numbers calculated for each models are connected by 
a thin dotted line. These numbers are in arbitrary units, but with the 
same normalization for all the cases.
}
\end{figure}

\begin{figure}
\centering
\psfig{file=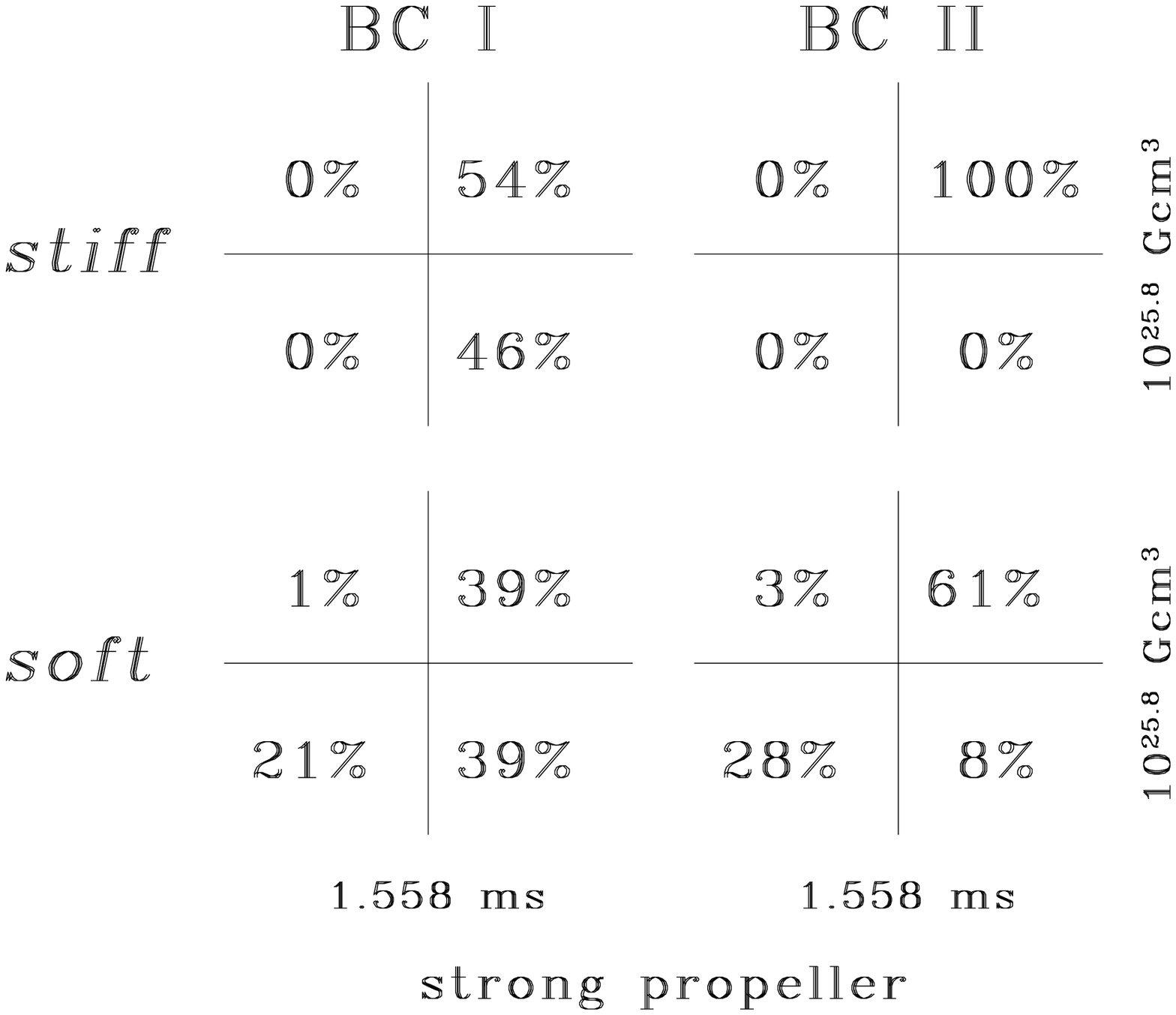,width=15cm}
\caption
{Distributions of the synthesized NSs, when a strong
propeller 
(${\cal F}_{que}=0.50; \Gamma_{que}=8$) is applied. Labels as in Figure~5.
}
\end{figure}

\begin{figure}
\centering
\psfig{file=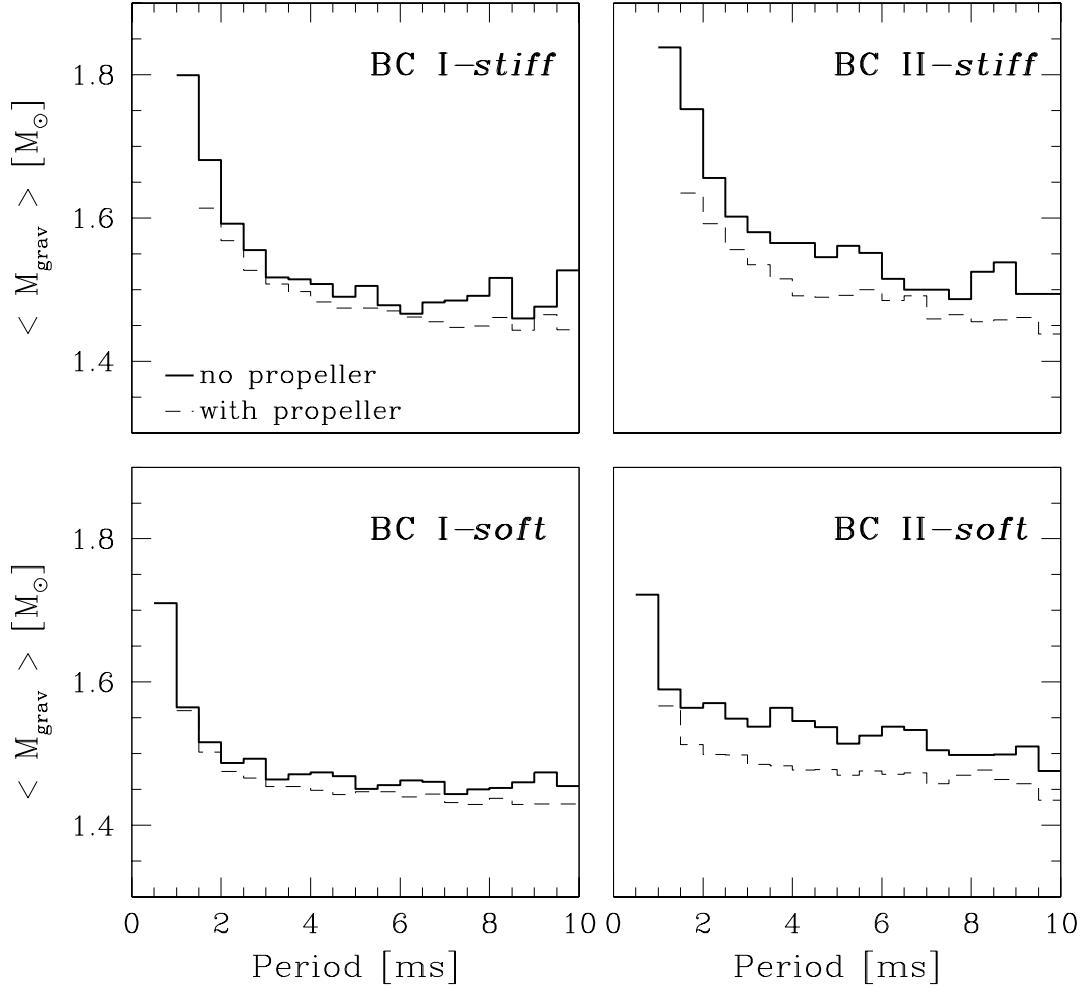,width=15cm}
\caption
{Average gravitational masses of the re-accelerated NSs 
as a function of their final spin period $P.$ $\mu$ is let vary over the whole 
range and the synthesized NSs are binned in $0.5$ ms wide intervals.
The initial mass of the static NSs is set equal to $1.40 {\rm M_{\odot}}$ for
all the cases. {\it Thick solid line} denotes the mass-distribution in 
absence of propeller, whereas {\it thin dashed line} gives the one with a 
strong propeller effect included $({\cal F}_{que}=0.50$ and $\Gamma_{que}=8).$
}
\end{figure}

\newpage

\begin{table}[tbh]
\begin{center}
{\large{\bf TAB. 1: Population syntheses parameters}}
\vskip 0.3truecm
\begin{tabular}{|c||c|c|c|}
{Physical quantity}         & {Distribution} &              {Values }                             &   {Units}
\\
\hline
\hline
 NS period at ${\rm t_{0}^{RLO}~(^*)}$  &     Flat     &                  1 $~~\to~~$ 100                      &   sec
\\
 NS $\mu$ at ${\rm t_{0}^{RLO}~(^*)}$  &   Gaussian   & Log$<\mu_0>$ = $~$28.50$~~;~\sigma$=0.32              &  G~cm$^{3}$
\\
${\dot{m}}$ in RLO phase$~(^{\sharp})$ &   Gaussian   & Log$<{\dot{m}}>$ =~ --~1.00$~;~\sigma$=0.50       & $\rm{{\dot M}_{E}}$
\\
 Minimum accreted mass                   &   One-value  &                       0.01                            & ${\rm {M_{\odot}}}$
\\
 RLO accretion phase time~($^\dag$)     &  Flat in Log &
 10$^{6}~~\to~~\tau_{RLO}^{max}$($^\ddag$) &    year
\\
 MSP phase time                          &  Flat in Log &         
 10$^{8}~~\to~~$3$~\times~$10$^{9}$           &    year
\\
\hline
\hline
\end{tabular}
\end{center}  
\begin{small}
\noindent
{\bf (*)~} ${\rm t_{0}^{RLO}}$ = initial time of the Roche Lobe Overflow phase
\\
{\bf ($\sharp$)~} baryonic accretion rate during the Roche Lobe Overflow phase
\\
{\bf ($\dag$)~} a Maximum accreted Mass of $0.5 M_\odot$ is permitted during the Roche Lobe Overflow phase
\\
{\bf ($\ddag$)~} maximum duration of the Roche Lobe Overflow phase; 
typical explored values: 5$\times$10$^7$ yr - 10$^8$ yr - 5$\times$10$^8$ yr
\end{small}
\end{table}

\newpage

\begin{table}[tbh]
\begin{center}
{\large {\bf TAB. 2: Population Syntheses Results}}
\vskip 0.3truecm
\begin{small}
\tabcolsep 0.06truecm
\begin{tabular}{c||c|ccc||c|ccc||c|c|}
\multicolumn{1}{c||}{} & \multicolumn{4}{c||}{\large {\bf BC I}} & \multicolumn{4}{c||}{\large {\bf BC II}} & \multicolumn{2}{c}{}   
\\  
\hline 
\hline
{\it EoS}     & {\it Number} & \multicolumn{3}{c||}{\it Percentage Ratio} & {\it Number} & \multicolumn{3}{c||}{\it Percentage Ratio} & 
{\it Kind of Evolution} & {\it Idx}
\\
~             & {\bf I} & {~~II/I~~} & {(II+III)/I} & {(III+IV)/I} & {\bf I} & {~~II/I~~} & {(II+III)/I} & {(III+IV)/I} & ~ & ~
\\
\hline
 ~            & {\bf 689} &  4.0   &  14.2 (11.5) &     63.5     & {\bf 823} &  0.3   &  0.3 (0.3)   &       0.0    & {\small standard}   &   i
\\
 ~            & {\bf 671} &  1.5   &   8.6 (6.9)  &     65.3     & {\bf 774} &  0.1   &  0.3 (0.3)   &       0.0    & {\small propeller}  &  ii
\\
{\large{\bf{\it Stiff}}}& {\bf 738} & 0.0 & 8.4 (2.3) &  110.3   & {\bf 1000} & 0.1   &  0.3 (0.3)   &       0.0    & {\small low $\dot m$+propeller}  & iii
\\ 
 ~            & {\bf 277}  & 0.0   &       0.0    &      0.0     & {\bf 275}  & 0.0   &  0.0         &       0.0    & {\small short$_{RLO}$+propeller} &  iv
\\
 ~            & {\bf 125}  & 0.0   &       0.0    &      0.0     & {\bf 116}  & 0.0   &  0.0         &       0.0    & {\small short$_{RLO}$+low $\dot m$+propeller} & v
\\
\hline
\hline
 ~            &  {\bf 742} & 10.1  &  91.9 (46.0) &    129.0     & {\bf 547}  & 19.4  &  74.3 (57.3) &      54.9    & {\small standard}  &  vi
\\
 ~            &  {\bf 730} &  8.0  &  83.9 (40.5) &    131.1     & {\bf 523}  & 16.3  &  73.8 (55.9) &      57.5    & {\small propeller} & vii
\\
{\large{\bf{\it Soft}}} & {\bf 720} & 0.6 & 85.1 (21.1) & 207.4  & {\bf 390}  & 12.1  &  42.5 (34.2) &      30.4    & {\small low $\dot m$+propeller} & viii
\\
 ~            &  {\bf 677} & 17.7  &  34.2 (34.2) &     17.2     & {\bf 640}  & 18.2  &  28.8 (28.8) &      10.6    & {\small short$_{RLO}$+propeller}  & ix
\\
 ~            &  {\bf 410} &  3.4  &   7.6 (7.6)  &      5.9     & {\bf 231}  &  7.0  &  11.9 (11.9) &       5.5    & {\small short$_{RLO}$+low $\dot m$+propeller} & x
\\
\hline
\hline
\end{tabular}
\end{small}
\end{center}
\begin{small}  
\noindent
{\bf I~=~} Number of synthesized objects in the first quadrant
($P>P_{min}$ and $\mu > \mu_{min}$). These numbers are 
normalized to the case (iii) for Boundary Condition BC II, 
set equal to 1000.
\\  
{\bf II/I~=~} Percentage ratio of the objects filling the II quadrant
over those filling the I quadrant.
\\
{\bf (II+III)/I~=~} Percentage ratio of the objects filling the II and the
III quadrant over those filling the I quadrant. In parenthesis the 
percentage ratio of the objects having $P<P_{min}$ and $\mu$ above
the death line over those filling the first quadrant.
\\
{\bf (III+IV)/I~=~} Percentage ratio of the objects filling the III and the
IV quadrant over those filling the I quadrant.
\\
{\bf Idx~=~} Index for the specific synthesized population. The different 
kinds of evolution are labelled as in the text.  
\end{small}
\end{table}

\end{document}